\documentclass{Interspeech2024}

\usepackage{array}
\usepackage{multirow}
\newcolumntype{C}{>{\centering\arraybackslash}p{1.475cm}} % centered column of 1.5cm width
\newcolumntype{L}{>{\raggedright\arraybackslash}p{1.475cm}} % left-aligned column of 1.5cm width
\newcolumntype{R}{>{\raggedleft\arraybackslash}p{1.475cm}} % right-aligned column of 1.5cm width

\interspeechcameraready

\title{Personalized Speech Enhancement Without a Separate Speaker Embedding Model}

\name[affiliation={}]{Tanel}{Pärnamaa}
\name[affiliation={}]{Ando}{Saabas}

\address{
  Microsoft Corporation}
\email{tanel.parnamaa@microsoft.com}


\begin{document}

\maketitle

\begin{abstract}
    
    Personalized speech enhancement (PSE) models can improve the audio quality of teleconferencing systems by adapting to the characteristics of a speaker's voice. However, most existing methods require a separate speaker embedding model to extract a vector representation of the speaker from enrollment audio, which adds complexity to the training and deployment process. We propose to use the internal representation of the PSE model itself as the speaker embedding, thereby avoiding the need for a separate model. We show that our approach performs equally well or better than the standard method of using a pre-trained speaker embedding model on noise suppression and echo cancellation tasks. Moreover, our approach surpasses the ICASSP 2023 Deep Noise Suppression Challenge winner by 0.15 in Mean Opinion Score.

\end{abstract}

\section{Introduction}

Audio signal enhancement components, such as noise suppression (NS), dereverberation and acoustic echo cancellation (AEC), are essential elements of modern teleconferencing systems. 
While they significantly improve speech signal quality, they have some limitations. For example, they are unable to remove unwanted human voices from the background, which are common in open office spaces or cafeteria environments. Additionally, echo cancellers typically fail to remove echoes resulting from delays that are too long or non-causal between the far end and microphone signals.
By knowing the user's voice, it is possible to enhance audio quality even in these challenging scenarios by isolating the user's voice from the input signal.

The standard approach to personalized speech enhancement (PSE) is based on the work on speaker-conditioned single-speaker extraction \cite{wang19h_interspeech, wang20z_interspeech, vzmolikova2019speakerbeam}. It consists of multiple stages. Firstly, a pre-trained speaker embedding model is used to extract an embedding vector from enrollment audio, representing the characteristics of a person's voice. Then, this  embedding is used in a separate speech extraction or enhancement model. The speaker embedding  model can be fixed, fine-tuned or trained jointly with the speech enhancement model.

Unfortunately, the multi-stage and multi-model approach has many practical difficulties: separate models need to be trained, deployed, maintained, and kept in synchronization, resulting in a significant engineering overhead, especially when deploying the models for edge devices.

In contrast, we propose an approach that does not rely on a separate embedding model. 
In particular, we note that for the speech enhancement model to make use of the speaker embedding, it needs to internally compute the representation of the speaker's voice in the input data that it is applied to, and compare it to the given embedding. Instead of having a separate speaker embedding model, we make use of this internal embedding to characterize a speaker's voice profile, and therefore only need a single model that is responsible for both speech enhancement and extraction of the speaker embedding. This change simplifies the training and deploying of personalized models.

This approach also offers other benefits, such as simplifying the enrollment process. In a typical personalized model usage scenario, the user must first provide an enrollment audio clip, for example, by reading a piece of text and recording it. Capturing enrollment audio automatically is more appealing because it removes the initial friction of using personalized models. However, speaker embedding models are usually too large and slow to run continuously on client devices. Running the speaker embedding model on a server might raise privacy concerns. By extracting the embedding directly from the speech enhancement model already in use for audio quality enhancement, we simplify the auto-enrollment process and minimize computational requirements because only one model is necessary.
 
We start with the state-of-the-art (SOTA) speech enhancement model, DeepVQE \cite{indenbom23_interspeech}. Firstly, we personalize it using the standard approach with a large pre-trained speaker embedding model. Next, we train the personalized model from scratch using its internal representation for speaker embedding, without changing its architecture or complexity. Our results show that this method matches the performance of the two-stage approach. Both models achieve SOTA results on the DNS Challenge noise suppression test data.

\section{Related work}

Recently, several challenges have been organized to benchmark methods for personalized NS and AEC. The top entries for the ICASSP 2022 and 2023 Deep Noise Suppression (DNS) Challenge personalized tracks utilized a two-stage approach with a separate embedding model for enrollment \cite{dubey2022icassp, dubey2023icassp, ju2022tea, ju2023tea, yan2023npu, yu2023tspeech}. Similarly, the winning entry for the ICASSP 2023 AEC Challenge used a separate embedding model \cite{cutler2023icassp, chen2023progressive}.

A common design choice in the two-stage approach is whether to use a fixed speaker embedding model, as done in VoiceFilter \cite{wang19h_interspeech} and PercepNet \cite{giri2021personalized}, or jointly learn or finetune the embedder, as done in SpeakerBeam \cite{vzmolikova2019speakerbeam} and SpEx \cite{xu2020spex}. Liu et al. showed in \cite{liu2023quantitative} that fine-tuning can improve audio quality on in-domain data, but the improvement does not generalize to out-of-domain data.
Moreover, none of the top participants in the DNS challenge finetuned the speaker embedding model, supporting the observation that transfer from simulated training data to real world test data is challenging with finetuned models.
This is unexpected because the representation learned for the speaker verification task, which is commonly used to train the embedding model, may not be optimal for the speech enhancement task. 

In \cite{liu2023quantitative}, the authors questioned the need for large speaker embedding models. They analysed the role of speaker embeddings in target speaker separation and found that the log-mel filterbank features worked surprisingly well on cross-dataset evaluation compared to learned features. However, they only evaluated on complete mixtures, and overlooked aspects like over-suppression and background speaker leakage, which are critical in teleconferencing systems.

Different to the standard approach, Sivaraman et al. used a method involving a mixture of local experts that does not require a reference speech utterance \cite{sivaraman2021zero}. During inference, a separate gating module embeds the audio and selects a specialised expert module that best fits the speaker. However, the training process involves a complex clustering based on speaker embeddings and multiple pre-training stages.

\section{Proposed method}

\begin{figure}[t]
  \centering
  \includegraphics[width=0.7\linewidth]{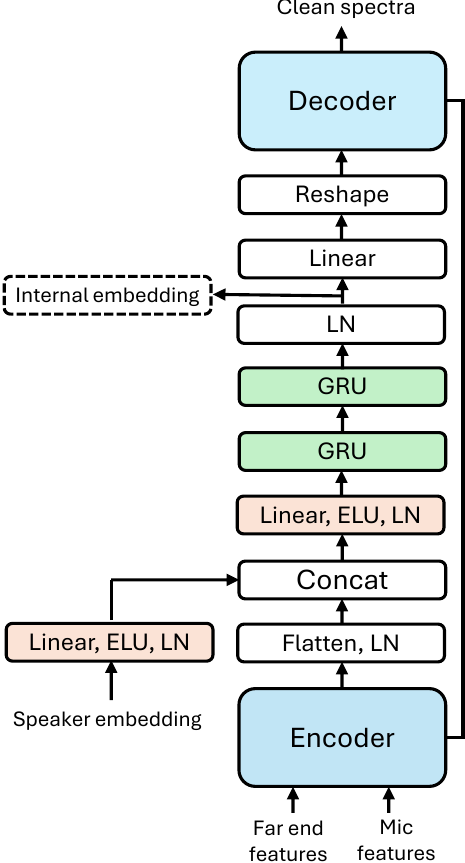}
  \caption{Model architecture with speaker information fusion. The figure shows how the speaker embedding is concatenated with the encoder features, the details of the temporal block, and the location of the internal embedding that we use to characterise speakers.}
  \label{fig:architecture}
\end{figure}

As a starting point, we take DeepVQE, a SOTA speech enhancement model for joint NS, AEC and dereverberation \cite{indenbom23_interspeech}.
This model shows best-in-class performance in the ICASSP 2023 AEC Challenge \cite{cutler2023icassp} and the non-personalized track of the ICASSP 2023 DNS Challenge \cite{dubey2023icassp}. 
The main building blocks of the model are an encoder, a Gated Recurrent Unit (GRU) bottleneck, a decoder, and a deep filter for output reconstruction called a complex convolving mask block. The input features are power law compressed complex spectra. The far end and microphone features are encoded separately, and then soft-aligned using a cross-attention mechanism. The decoder uses sub-pixel convolutions for upsampling.  For more details, we refer the reader to \cite{indenbom23_interspeech}.  We will condition the model based on speaker embedding to make it personalized, and show how to use its internal embedding to describe enrollment audio.

\subsection{Personalized DeepVQE}

To make the model personalized and suppress neighboring talkers, we need to fuse speaker information to the model. We first describe the standard two-stage approach to personalization using a separate pre-trained embedding model. In Section \ref{internal_emb}, we show how we can remove the separate embedding model, without requiring any changes to the speech enhancement models.

In the two-stage approach, a speaker embedding vector is extracted from enrollment audio using a pre-trained model, then linearly projected to match the size of flattened features before the GRU layer, followed by an activation function and layer normalization (LN) \cite{ba2016layer}. The output is then concatenated with the flattened features, and linearly transformed to match the original size of the flattened features, which is again followed by an activation function and LN. The fusion is shown in Figure \ref{fig:architecture}.

We use a different residual block and temporal block compared to the DeepVQE architecture. Specifically, we adopt an inverted residual block \cite{howard2019searching}, but drop the squeeze and excitation block \cite{hu2018squeeze}, and use standard convolutions instead of depthwise convolutions. Moreover, we use two GRU layers, and layer-normalize the input of the first GRU layer and the output of the last GRU layer. The temporal block is shown in Figure \ref{fig:architecture}.
These changes showed improvement over the baseline DeepVQE architecture in the PSE scenario.

\subsection{Proposed embedding extraction}
\label{internal_emb}

To achieve personalization, the PSE model has to represent speaker information in its internal states. Intuitively, when the input signal contains a single voice, the PSE model compares its representation of the current speaker with the enrollment embedding to decide whether to suppress the speech. How can this speaker information be extracted from the PSE model? Most speech enhancement models employ an encoder-decoder approach with a temporal block in between. A natural place to extract an internal embedding is from the temporal component. For example, models like DeepVQE, PercepNet, VoiceFilter-Lite, and E3Net \cite{eskimez2022personalized} employ an RNN cell, from which the internal embedding can be extracted. In transformer-based models such as MTFAA-Net \cite{zhang2022multi}, the internal embedding can be obtained after a specified transformer layer. 

In our experiments, we extract the internal embedding after the temporal block, and average the frame-based embeddings gathered from enrollment audio. Specifically, we extract $T$ frames of features from enrollment audio, which produce $T$ embeddings of size $K$. We then average these to get a single embedding of size $K$. This embedding may also contain other information, such as the presence of echo or noise categories, but the PSE model will learn to use information from the state that is conducive for minimizing the loss. 

The internal embedding could also be gathered from other locations in the PSE model, such as before the temporal block or between the GRU layers. Moreover, when taking the internal embedding after the fusion of speaker information, we also need to specify the initial speaker embedding when mapping enrollment audio to a speaker embedding. We use a vector of zeros for that.

We want to emphasize that this change, moving the embedding extraction from a separate model to the speech enhancement model itself, does not require any change in the speech enhancement model. Both the architecture and the size of the enhancement model remain the same in both approaches.

\section{Experimental setup}

{ % start of scope
\setlength{\tabcolsep}{1pt}
\centering
\begin{table*}[h]
\caption{Objective quality comparison with baselines for AEC in far end single talk and double talk scenarios, personalized NS, target speaker over-suppression, and background speakers removal quality. All models use the PVQE-S architecture except the Noisy baseline.}
%\begin{tabular}{lrrrrrrrrrr}
\begin{tabular}{L|CC|CC|CCC|C|C|C}
%\toprule
          Speaker embedding &  AECMOS Echo ↑ &  ERLE ↑ &  AECMOS Echo ↑ &  AECMOS Deg ↑ &  SIG ↑ &  BAK ↑ &  OVRL ↑ &  PESQ ↑ &  TSOS ↓ &  BAK SUPPR ↑ \\
\midrule
    \textit{(Noisy)} &           1.99 &     0.0 &           1.79 &          3.85 &   4.16 &   2.27 &    2.65 &    1.61 &   \textbf{0.000} &         0.00 \\
\midrule
      Empty &           4.58 &    68.9 &           4.55 &          \textbf{3.98} &   4.03 &   3.97 &    3.53 &    2.68 &   0.003 &         5.37 \\
          FBANK &           4.63 &    74.7 &           4.56 &          3.96 &   4.03 &   3.97 &    3.53 &    2.66 &   0.015 &        16.96 \\
Separate &           4.64 &    74.9 &           4.57 &          3.97 &   4.02 &   4.04 &    3.56 &    \textbf{2.72} &   0.027 &        \textbf{34.30} \\
Internal &           \textbf{4.68} &    \textbf{75.7} &           \textbf{4.61} &          3.92 &   4.01 &   \textbf{4.13} &    \textbf{3.59} &    2.67 &   0.010 &        33.04 \\
%\bottomrule
\end{tabular}

\label{tab:comparison}
\end{table*}
} % end of scope

{ % start of scope
\setlength{\tabcolsep}{2.75pt}
\centering
\begin{table*}
\caption{Subjective MOS results on the DNS challenge blind test set. We include the challenge winner and personalized baseline. The topmost three models are larger, and the rest are smaller models.}
\newcolumntype{C}{>{\centering\arraybackslash}p{1.18cm}} % centered column of 
\begin{tabular}{l|l|CCC|CCC|C|C}

  \multicolumn{1}{l}{\phantom{}} & \multicolumn{1}{l}{\phantom{}}  &\multicolumn{3}{|c}{Track 1 - Headset} & \multicolumn{3}{|c|}{Track 2 - Speakerphone} & \multicolumn{2}{r}{} \\
 \midrule
Model& Embedding & SIG & BAK & OVRL & SIG & BAK & OVRL & Avg & CI \\
 \midrule
PVQE-L &  Internal  & \textbf{3.61} & \textbf{3.10} & \textbf{2.96} & \textbf{3.70} & \textbf{3.12} & \textbf{3.00} & \textbf{3.25} & 0.04 \\
PVQE-L &Separate & 3.58 & 3.03 & 2.89 & \textbf{3.70} & 3.09 & 2.95 & 3.21 & 0.04 \\
Challenge winner \cite{ju2023tea} &Separate & 3.58 & 2.87 & 2.75 & 3.69 & 2.90 & 2.83 & 3.10 & 0.04 \\
\midrule
PVQE-S & Internal & \textbf{3.47} & \textbf{2.70} & \textbf{2.58} & \textbf{3.59} &\textbf{2.76} & \textbf{2.68} & \textbf{2.96} & 0.04 \\
PVQE-S & Separate & \textbf{3.47} & 2.64 & 2.52 & 3.48 & 2.60 & 2.54 & 2.88 & 0.04 \\
Challenge baseline (E3Net \cite{eskimez2022personalized})&Separate & 3.28 & 2.62 & 2.46 & 3.50 & \textbf{2.76} & 2.64 & 2.87 & 0.04 \\
\midrule
Noisy & -  & 3.62 & 1.28 & 1.29 & 3.71 & 1.25 & 1.25 & 2.06 & 0.02 \\
%\bottomrule
\end{tabular}

\label{tab:subjective_comparison}
\end{table*}
} % end of scope

To understand the effectiveness of the proposed method, we compare it to multiple baseline models. Firstly, we use a model without a speaker embedding, representing a scenario where the closest speaker to the microphone is extracted from the audio signal. This condition helps to assess the overall usefulness of speaker information. It is implemented by setting the speaker embedding to zeros in the personalized DeepVQE model. Secondly, we use log-mel filterbank features as the speaker embedding to understand the improvement of learned features over simple feature extraction. For this, we follow \cite{liu2023quantitative} and compute 80-dim FBANK features, and concatenate the temporal mean and standard deviation to get an embedding of size 160. Thirdly, we use a Res2Net model trained for speaker verification to extract speaker features \cite{zhou2021resnext}.

\subsection{Training data}

We adopt the data generation approach outlined in \cite{indenbom23_interspeech}, with modifications to include enrollment clips and utilize background speech in addition to background noises. Beyond the datasets provided in the ICASSP 2023 AEC and DNS challenges, we also use VoxCeleb2 \cite{chung2018voxceleb2}, which we pre-process with a noise suppressor to remove background noises. As the speaker identity information can be wrong or clips might contain multiple speakers, we remove clips where intra-speaker distance of clip embeddings is high \cite{wang2023framework}. The sampled enrollment clips are 10 seconds long, and 50\% of them are noisy with an SNR sampled from the range [0, 40]. The training clips are 40 seconds in length, and 30\% of the clips contain background speech with an SIR sampled from the range [0, 20].

\subsection{Evaluation data}

For evaluation, we use the AEC Challenge 2023 blind test set to assess AEC performance, DNS Challenge 2023 blind test set to evaluate personalized NS performance, AMI dataset \cite{carletta2005ami} for evaluating target speaker over-suppression. The AEC and DNS test set are used as provided in the challenges. From the AMI dataset, we sampled 1549 clips where only a single speaker is present, and ensured enrollment audio was recorded on a different device. This is important because most over-suppression cases happen when the target speech differs from the enrollment, such as when the audio is recorded from a different location or using a different microphone, which are common scenarios in practice. The average test clip length is 23.8 seconds, and the average enrollment length is 12.2 seconds. For the evaluation of background speech suppression, we sampled clips from LibriVox data such that the enrollment speaker and nearend speaker are different. The perfect outcome is to suppress all content in such clips. Additionally, we use internally collected clips where the target speaker is present at the beginning, followed by a long period of background talk. We use 220 clips in total for background speech suppression evaluation, with an average clip length of 42.3 seconds and an average enrollment length of 12.6 seconds.
Furthermore, we created 200 synthetic 15-second mixtures based on LibriVox data to evaluate target speaker extraction with reference-based metrics.

\subsection{Hyperparameters}

We use a 20ms squared root Hann window, a hop size of 10ms, a discrete Fourier transform length of 320, and sample audio at 16 kHz. We extract power law compressed complex spectra from the noisy microphone input and far end signal. 
If the far end signal is missing, we set the signal to zero.

We adapt the DeepVQE-S configuration for comparison study, and name the personalized configuration as PVQE-S.
We use two encoder blocks for the far end and microphone signal, followed by an alignment block with a history of 100 frames to align the features in time with a maximum delay of 1 second. The far end branch uses 8 and 24 filters, and the microphone branch uses 16 and 40 filters. The microphone features and aligned far end features are concatenated, and fed into a combined encoder that consists of two blocks of 56 and 24 filters. The kernel size for the encoder is 2x3, with the first dimension representing temporal and the second dimension frequency axis.
For the temporal block, we use two GRU layers of size 256 followed by a linear projection. Moreover, we layer-normalize the input and output of the temporal block.
The decoder uses sub-pixel convolutions and has 4 blocks of 40, 32, 32, and 27 filters.
The combined encoder and the first two decoder blocks make use of a residual block with an expansion factor of 0.7 \cite{howard2019searching}. Exponential linear unit (ELU) is used as an activation function throughout the model \cite{clevert2015fast}.
No look-ahead is used in the model.

The speaker embedding is fused to the model before the temporal block. Specifically, a linear layer transforms the speaker embedding to a size of 240, followed by an activation function and LN. The output is then concatenated with the flattened features from the encoder and projected back to the size of the flattened features. In experiments with a pre-trained speaker embedding model, we use a Res2Net-based speaker verification model \cite{zhou2021resnext}, which returns an embedding of size 128 and has 15M parameters. In the proposed method, we extract the internal embedding from the output of LN which comes after the last GRU layer.

The models are trained using the Adam optimizer with a learning rate of $6\times10^{-5}$ and a weight decay of $1\times10^{-7}$ for 234k iterations on batches of 64 samples. A complex compressed MSE loss \cite{braun2021consolidated} is used with an exponent of 0.3 and a beta of 0.7.

\section{Results}

\subsection{Objective evaluation}

For objective evaluation, we use non-intrusive neural network-based mean opinion score (MOS) estimators, personalized DNSMOS P.835 \cite{reddy2022dnsmos} and AECMOS \cite{purin2022aecmos}, to assess the quality of noise and echo removal.
Personalized DNSMOS P.835 provides three scores: speech quality (SIG), background noise quality (BAK), and overall quality (OVRL) of the audio. AECMOS gives two scores: echo removal (AECMOS Echo) and signal degradation (AECMOS Deg) quality. Additionally, we report echo return loss enhancement (ERLE) for the far end single talk scenario and Perceptual Evaluation of Speech Quality (PESQ) \cite{rix2001perceptual} for simulated two-speaker mixtures. 
Furthermore, we report the target speaker over-suppression metric (TSOS) from \cite{eskimez2022personalized}. 
Finally, to evaluate the effectiveness of the models in scenarios where only interfering speakers are present, we measure the signal energy reduction in decibels, where a higher number indicates better performance~(BAK SUPPR).

Table \ref{tab:comparison} shows the objective metrics for original noisy data, baseline methods and proposed models. Firstly, we can see that the use of speaker embedding gives a clear performance boost in background speech suppression scenarios. This is indicated by the BAK and BAK SUPPR scores, where models using the embedding outperform the no-embedding and filter-bank based approaches, without degrading signal scores.

At the same time, we see that the model using an internal embedding gives similar results to the two-stage model, even outperforming it for some metrics. Especially noteworthy is that while the internal embedding model gives as good or better results for background speech removal, it does not do so at the cost of near-end over-suppression. In fact, the opposite is true, the TSOS metric shows 2.7x reduction in over-suppressed frames.

\subsection{Subjective evaluation results}

To evaluate our approach with respect to the state of the art, we compare it against the baseline and winning models of the 2023 DNS Challenge. For a fair comparison, we increased the model size while maintaining the real-time constraints specified by the challenge rules. Specifically, we increased the layer sizes to match the base configuration in \cite{indenbom23_interspeech} and trained the large models on super-wideband data. 

We conducted a subjective evaluation on the DNS challenge data using  the personalized version of ITU-T P.835 framework \cite{naderi_subjective_2021}, which uses 5 seconds of clean enrollment speech from primary talkers to help human raters recognize the primary speaker’s voice when scoring the clips. The evaluation was run on a crowd-sourcing platform with 10 raters per clip.

Our large models (PVQE-L) outperformed the challenge winner, achieving a higher BAK score of over 0.2 MOS on both tracks without compromising signal quality. For large models, both the pre-trained and the internal speaker embedding extraction yielded similar results. However, for small models, the internal embedding method was more effective, improving the track 2 overall score by 0.14. This demonstrates that using an internal embedding is especially useful for small real-time models, leading to a simple and effective approach for balancing the representations given by a speaker embedding model and the PSE model.

\subsection{Inference speed}
We measure the inference speed of the large and small model on an Intel Core i7 10700K@3.8GHz CPU. We use a single-thread configuration and report the average inference time over 100,000 frames. The large model has 8.38M parameters and takes 3.64ms per frame, while the small model has 1.07M parameters and takes only 0.135ms per frame.
This means that the small model can process audio signals 74 times faster than real-time, achieving a real-time factor of 0.0135. The small model demonstrates a remarkable balance between performance and complexity, making it suitable for real-time teleconferencing applications.

\section{Conclusions}

In this paper, we proposed a novel and simple approach to personalized speech enhancement that does not require a separate speaker embedding model. We showed that the internal representation of the speech enhancement model can be used as the speaker embedding. We evaluated our approach on two speech enhancement tasks: noise suppression and echo cancellation, and showed that the model achieves state-of-the-art performance on the DNS Challenge data. We compared our approach with the standard method of using a pre-trained speaker embedding model, and found that our approach improves background noise quality for small models. 

\bibliographystyle{IEEEtran}
\bibliography{mybib}

\end{document}